\documentclass[aps,prb,floatfix,showpacs]{revtex4-1}
\usepackage{graphicx}
\begin{document}
\title{Topological insulators in the quaternary chalcogenide compounds and
ternary famatinite compounds}
\author{Y. J. Wang$^1$, H. Lin$^1$, Tanmoy Das$^{1,2}$, M. Z. Hasan$^{3}$, and A. Bansil$^1$ }

\address{ $^1$Department of Physics, Northeastern University,
 Boston, MA 02115, USA\\
 $^2$Theoretical Division, Los Alamos National Laboratory, Los Alamos, NM 87545, USA\\
 $^3$ Joseph Henry Laboratories of Physics, Princeton University, Princeton,
New Jersey 08544, USA\\}


\begin{abstract}
We present first-principles calculations to predict several three
dimensional (3D) topological insulators in quaternary
chalcogenide compounds which are made of I$_2$-II-IV-VI$_4$ compositions and
in ternary compositions of I$_3$-V-VI$_4$
famatinite compounds. Among the large members of these two families, we give
examples of naturally occurring compounds
which are mainly Cu-based chalcogenides. We show that these materials are
candidates of 3D topological insulators or can
be tuned to obtain topological phase transition by manipulating the atomic
number of the other cation and anion elements.
A band inversion can occur at a single point $\Gamma$ with considerably
large inversion strength, in addition to the
opening of a bulk band gap throughout the Brillouin zone. We also
demonstrate that both of these families are related to
each other by cross-substitutions of cations in the underlying tetragonal
structure and that one can suitably tune their
topological properties in a desired manner.
\end{abstract}
\maketitle

\section{Introduction}

The recent discovery of topological insulators \cite{review,reviewMZCL,
Kane1st,FuKane,DavidNat1,KonigSci,DavidTunable,TIbasic,BernevigSciHgTe,Roy,MooreandBal}
has realized a long-sought opportunity of implementing exotic quantum
phenomenon in practical materials. \cite{Majorana, ZhangDyon,
KaneSCproximity, KaneDevice, cenke, palee, dhlee} However, existing
topological materials are restricted to binary and ternary compositions of
heavy
elements\cite{DavidNat1,KonigSci,DavidTunable,DavidScience,MatthewNatPhys,Noh,ZhangPred,ChenBiTe,BiTeSbTe,
Roushan, WrayCuBiSe,heuslerhasan,TlBiTe2}, which limits the possibility of
combining correlated electronic, magnetic, superconducting and other local
order properties with topological phenomenon. It is highly desirable,
therefore, to expand the menu of available materials into the domain of
compounds which are relevant for device fabrication with tuning
capabilities and accessible crystalline cleavage.

Topological insulator order and local order of spontaneous symmetry
breaking such as magnetism and superconductivity has recently been
classified in different spatial dimensions and with different discrete
symmetries. The band topology in the two-dimensional quantum spin-Hall
effect or three dimensional non-trivial topological insulators are
required to possess a time-reversal invariant number $Z_{2}$ = -1.
Materials with strong spin-orbit coupling can host non-trivial topological
phases in both two and three spatial dimensions when the time-reversal
symmetry remains invariant. The topological insulating materials realized
to date are based on the binary and ternary combinations of mostly heavy
metals ranging from HgTe,\cite{KonigSci} Bi/Sb-based binary
systems\cite{DavidNat1,DavidTunable,DavidScience,MatthewNatPhys, Noh,
ZhangPred,ChenBiTe,BiTeSbTe, Roushan, WrayCuBiSe} to Tl-based ternary
compounds\cite{TlBiTe2}. However, the
principle allure of topological insulators stems from their potential
technological applications, which require the time-reversal-invariant
topological ground state to be placed along with the broken symmetry of
local orders.\cite{Majorana, ZhangDyon,KaneSCproximity, KaneDevice, cenke,
palee, dhlee} Of particular interest are the magnetic atom doped
topological insulators for realizing effects of magnetic impurities and
ferromagnetism on the topological surface
states.\cite{Liumagimp,Hormagimp,Yumagimp} Furthermore, the realization of
elementary excitations that satisfy non-Abelian statistics $-$ the
so-called Majorana fermions, important for topological quantum computers,
requires proximity of superconducting and ferromagnetic insulating
phases.\cite{KaneDevice,Akhmerov,Tanaka} Therefore, it is important to
search for material classes which provide more chemical and structural
flexibility to tune the ground state properties using cross-substitution
of elements.

Here, we report the theoretical prediction of three dimensional topological
insulator in the pristine state of existing materials in
quaternary chalcogenide compounds which are made of I$_2$-II-IV-VI$_4$
compositions and in ternary compositions of
I$_3$-V-VI$_4$ famatinite compounds. Both families are related by
cross-substitutions of cations in the underlying
tetragonal structure. Quaternary semiconductors, especially, exhibit more
flexible properties arising from their enhanced
chemical and structural freedom and provide a handle on the surface quantum
control through the large possibility of substitution
with the magnetic, non-magnetic and other correlated electronic elements.
Through the ternary-to-quaternary mutation and control of
the atomic configuration, one can perform band engineering to tailor the
topological quantum phenomenon for transport and
quantum information processing application.

We begin the discussion with the ternary composition of I$_3$-V-VI$_4$
compounds which are known to crystallize in
famanitine structures and are basically of the sphalerite-type. These
materials belong to the space group
$I{\bar 4}2m$, in which the group VI atoms are surrounded by three group I
and one group V elements. As a result,
they obey the octet rule and form a (I-VI)$_3$(V-VI) superlattice structure,
see Fig.~1(a) for Cu$_3$SbSe$_4$ as an example.
In contrast to the high-symmetry zinc-blende sublattice of ternary
compounds\cite{TlBiTe2,Satotl,Chentl}, famatinite
structure naturally achieves a tetragonal lattice distortion along the
$c-$axis ($c<2a$) due to the turning on of the strong
interlayer coupling between the two cation planes. This also results in a
mismatch between the cation-anion bond lengths in two
zinc-blende formula units of the unit cell which helps lower the total
energy of the famatinite ground state phase.

\begin{figure}
\centering
\includegraphics[width=8.5cm]{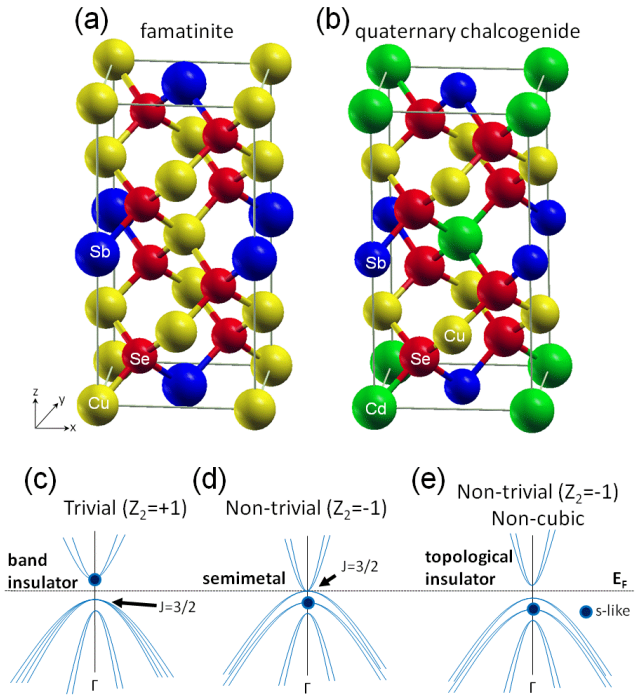}
\caption{\label{fig:sketch} {\bf Crystal structure and the property of the
topological band-inversion.}
(a), The crystal structure of ternary famatinite compounds is given for
a representative material Cu$_{3}$SbSe$_{4}$.
(b) The crystal structure of quaternary chalcogenide compounds is
illustrated for Cu$_2$CdSbSe$_4$ as an example.
(c)-(e) Band structures illustrated schematically near the time-reversal
invariant point $\Gamma$ for
a trivial band insulator in (c), a non-trivial topological semimetal in
(d) and a non-trivial
topological insulator in (e). Dark-blue dots represent the twofold
degenerate $s-$like states.
In the non-trivial case, band inversion occurs when s-like states drops
below the four-fold degenerate $p-$like states $(J=3/2)$ (blue lines).
Owing to the lattice distortion, the degeneracy of the $J=3/2$ states
vanishes in the present non-cubic crystal structures.}
\end{figure}
Famatinite compounds I$_3$-V-VI$_4$ evolve to the quaternary chalcogenides
I$_2$-II-IV-VI$_4$, when one of the group-I
element of the former compounds is replaced by one group-II element
and the group-V element is changed to group-IV
element; see Fig.~1(b) for example Cu$_2$CdSbSe$_4$. High-resolution
transmission electron microscopy and X-ray
diffraction analysis shows that these materials have tetragonal
crystallographic structure which belongs
to $I{\bar 42m}$ space group. The structure can be described by a
(I-II)$_2$(II-VI)(IV-VI) sublattice of the two zinc-blende
formula units as before.\cite{Chenquaternary} The substitution of larger
atoms in the tetragonal basis of these systems helps
expand the lattice and thus both the tetragonal distortion and the
crystal-field splitting increases considerably. Although
these changes are relatively small, nevertheless, quaternary chalcogenides
have more structural freedom and more complicated
electronic and chemical properties than their binary or ternary
counterparts.

The typical topological phases can be classified as trivial band insulator,
a non-trivial topological semimetal (zero-bandgap)
and a non-trivial topological insulator(finite-bandgap). The corresponding
band symmetries near $\Gamma$ point are expressed
schematically in Fig.~1{c-e} respectively. For a trivial topological phase
with the $Z_2$ topological order equal to +1 in a
zinc-blende crystal such as CdTe, a bandgap exists at the Fermi-level as
well as the twofold degenerate $s-$like states marked
by darked-blue dot lies in the conduction band. In a non-trivial topological
phase case such as HgTe, band inversion occurs
whereas $s-$like states drop below the fourfold degenerate $p-$like states
of $J=3/2$ multiplet, with $Z_2$=-1. The fourfold
degeneracy is due to the cubic symmetry in the zinc-blende structure. To
achieve a topological insulating state, the degeneracy
is required to be lifted and a band gap is opened (Fig.1(e)). Since the
non-cubic famatinites and quaternary chalcogenides are
derived from zinc-blende structures, the non-trivial topological insulating
phase could be realized in these materials. Indeed,
our first-principles calculations predict that 3D non-trivial
topological insulating phase exists in the compounds of famatinite
and quaternary chalcogenides families.


\section{Methods}
The crystal structures of ternary famatinite compounds and quaternary
chalcogenide compounds are taken from literature
\cite{Liter1,Liter2,Liter3,Liter4,Liter5,Liter6,Liter7,Liter8,Liter9,Liter10,Liter11,Liter12,Liter13,Liter14,Liter15}.
First-principles band calculations were performed with the linear
augmented-plane-wave (LAPW) method using the WIEN2K
package \cite{wien2k} in the framework of density functional theory (DFT).
The generalized gradient approximation (GGA)
of Perdew, Burke, and Ernzerhof \cite{PBE96} was used to describe the
exchange-correlation potential. Spin orbital coupling
(SOC) was included as a second variational step using a basis of
scalar-relativistic eigenfunctions.

\section{Results}
The bulk band structures along high symmetry path
$M(\pi,\pi)-\Gamma(0,0)-X(\pi,0)$ are shown in Fig.~2 for two representative
compunds
Cu$_3$SbS$_4$ and Cu$_3$SbSe$_4$ which belong to the famatinite family. The
effect of the zone-folding is clearly evident at the high-symmetry
point $\Gamma$. In
the ground state of these compounds, the structural compression
along c-axis (i.e. $c<2a$) compared to cubic zinc-blende lattice lifts the
four-fold degeneracy of the cation $p-$states at the $\Gamma-$point
owing to the crystal-field splitting. In Cu$_3$SbS$_4$, an insulating energy
gap between the $p-$states is present at the Fermi level throughout
the Brillouin zone as shown in Fig.~2(a). However, as the nuclear charge of
the anion is increased from S atom to Se atom, such substitution
causes the conduction band to drop below the Fermi level at the X-point,
yielding an electron pocket, see Fig.~2(b). Simultaneously, the valence
band maximum also gradually moves above the Fermi-level making
Cu$_3$SbSe$_4$ a metal. Nevertheless, a finite direct gap persists
throughout the
Brillouin zone and thus the $Z_2$ topological invariant can still be defined
for the valence bands in these two materials and the inverted band
order is retained at only $\Gamma-$points. Therefore, the Cu$_3$SbS$_4$ and
Cu$_3$SbSe$_4$ are topologically non-trivial insulator and metal,
respectively.

\begin{figure}
\centering
\includegraphics[width=5.5cm]{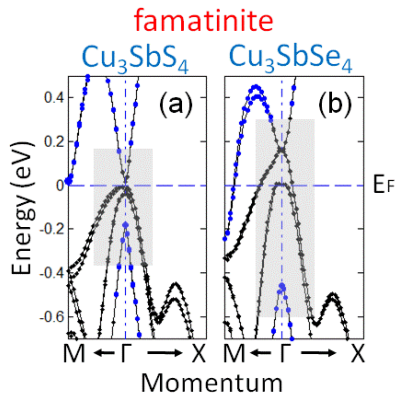}
\caption{\label{fig:bulkbands2} {\bf The bulk band dispersion of famainite
compounds with non-trivial topological phases.}
(a)-(b) Electronic structure of Cu$_3$SbS$_4$ and Cu$_3$SbSe$_4$. In both
figures, the black dots are the $p-$states
of the cation sites while the size of the blue dots is proportional to the
probability of $s-$orbital occupation on the anion
site. The gray shadings highlight the low-energy region near $\Gamma-$point
where the inverted band-inversion has occurred.
}
\end{figure}
The mutation of the crystal structure from the ternary to a quaternary
compound is also evident in the electronic structure of these two families
of
materials as the overall changes in the electronic structure are only subtle
near the energy region of present interests. Due to the tetragonal symmetry,
all the Cu-based quaternary materials listed in Fig.~3 host a gap in the
$p-$states close to the Fermi-level, where the gap magnitude depends on the
extra cation cross-substitutions. With increasing atomic mass from Zn to Cd
to Hg atom, this gap decreases mainly due to the downshift of the $s-$like
conduction bands at $\Gamma-$point. The Fermi-level goes through the energy
gap of the $p-$states in most materials, harboring a topological insulating
state in these compounds. However, only in two compounds Cu$_2$ZnGeTe$_4$
and Cu$_2$CdGeTe$_4$, the low-energy electronic properties reveal that the
conduction band minimum is shifted to the $X$ point and drops below the
Fermi level to form bulk electron pockets. Simultaneously, the
concave-upward
shaped valence band maximum pops up above the Fermi level at $\Gamma-$point,
making these systems intrinsically bulk metallic. We find very similar
nature of the band-inversion across these materials in which the split
$p-$states lie in energy above the twofold-degenerate $s-$ states at a
single
time-reversal momentum $\Gamma-$point, representing a band-inversion
relative to the natural order of $s-$and $p-$type orbital derived band
structure. Therefore, these materials are intrinsically bulk non-trivial
topological insulators or semimetals, except
Cu$_2$ZnGeS$_4$, Cu$_2$ZnSnS$_4$, and Cu$_2$HgGeS$_4$ which are trivial
topological insulators.
\begin{figure}
\centering
\includegraphics[width=12cm]{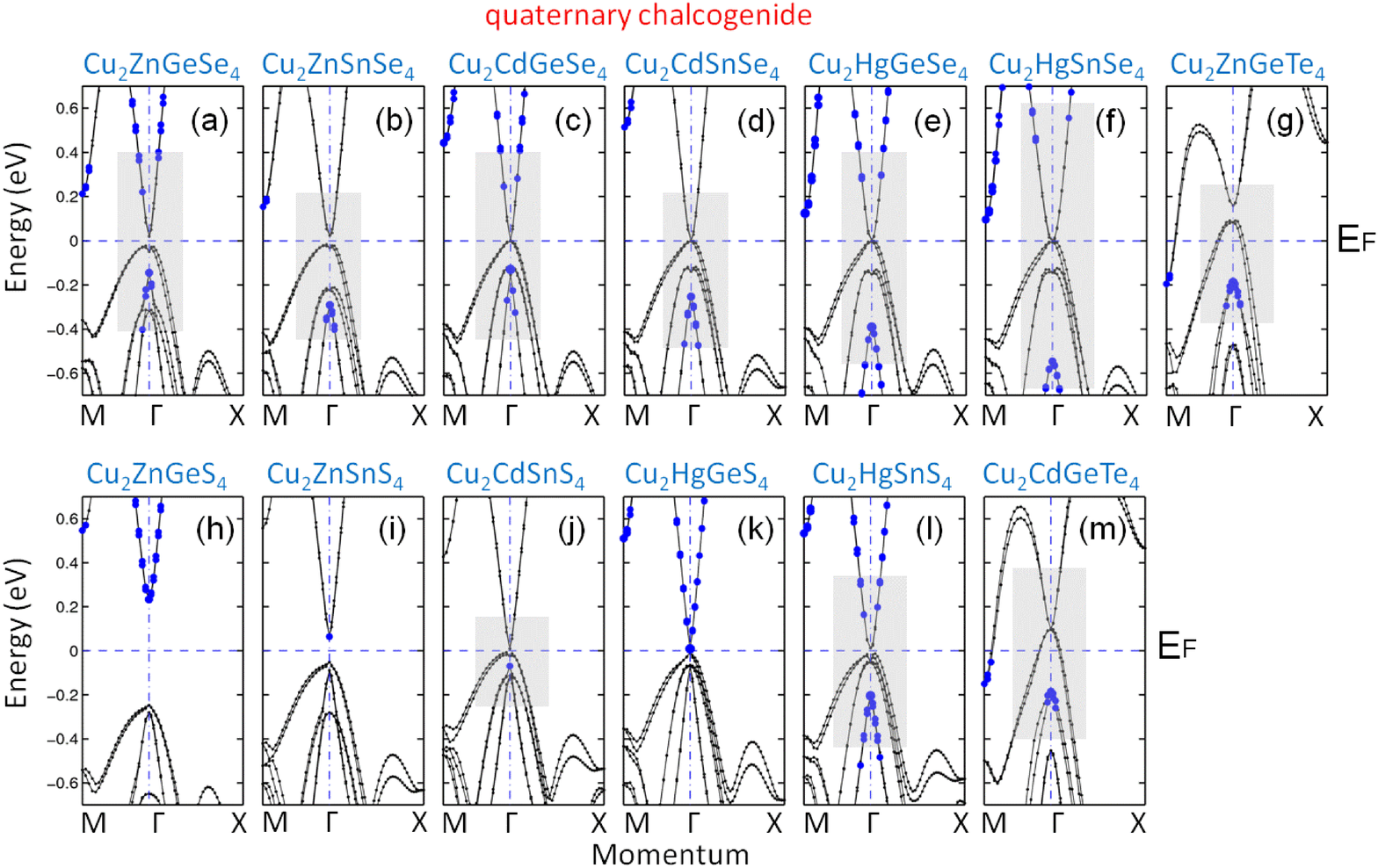}
\caption{\label{fig:bulkbands3} {\bf Electronic structure of quaternary
chalcogenide compounds.}
Bulk electronic structure of several Cu-based quaternary chalcogenides
compounds as indicated in the title of each
figures. The black dots, blue dots and the gray shading have the same
meaning as in Fig.~2. The compounds without a
gray-shaded area are trivial topological insulators without any inverted
band symmetries whereas the others are
non-trivial topological insulators or semimetals with $Z_2=-1$ topological
order.}
\end{figure}

Finally, we demonstrate our tuning procedure to generate a topological phase
transition from a trivial band insulator to the non-trivial topological
insulator. In the present study, we take advantage of the structural freedom
of the solids in ternary famatinite and quaternary chalcogenides to achieve
non-trivial topological phases. In the Fig.~4, we discuss one representative
example of this route here in going from the quaternary chalcogenide
Cu$_{2}$ZnSnS$_{4}$ as the starting point to the ternary famatinite
Cu$_3$SbS$_4$ compound. Note that the topological phase transition is a
generic
feature and there exists numerous such route. Cu$_{2}$ZnSnS$_{4}$ is a
trivial band insulator whereas Cu$_3$SbS$_4$ is a topological insulator in
their pristine conditions as discussed before in Figs.~3(i) and 2(b),
respectively. We systematically manipulate the overall crystal structure of
our starting compound by making the atomic number $Z$ of the constituent
atoms, the lattice constant and the internal displacement of the anion as
variable parameters.

\begin{figure}
\centering
\includegraphics[width=12cm]{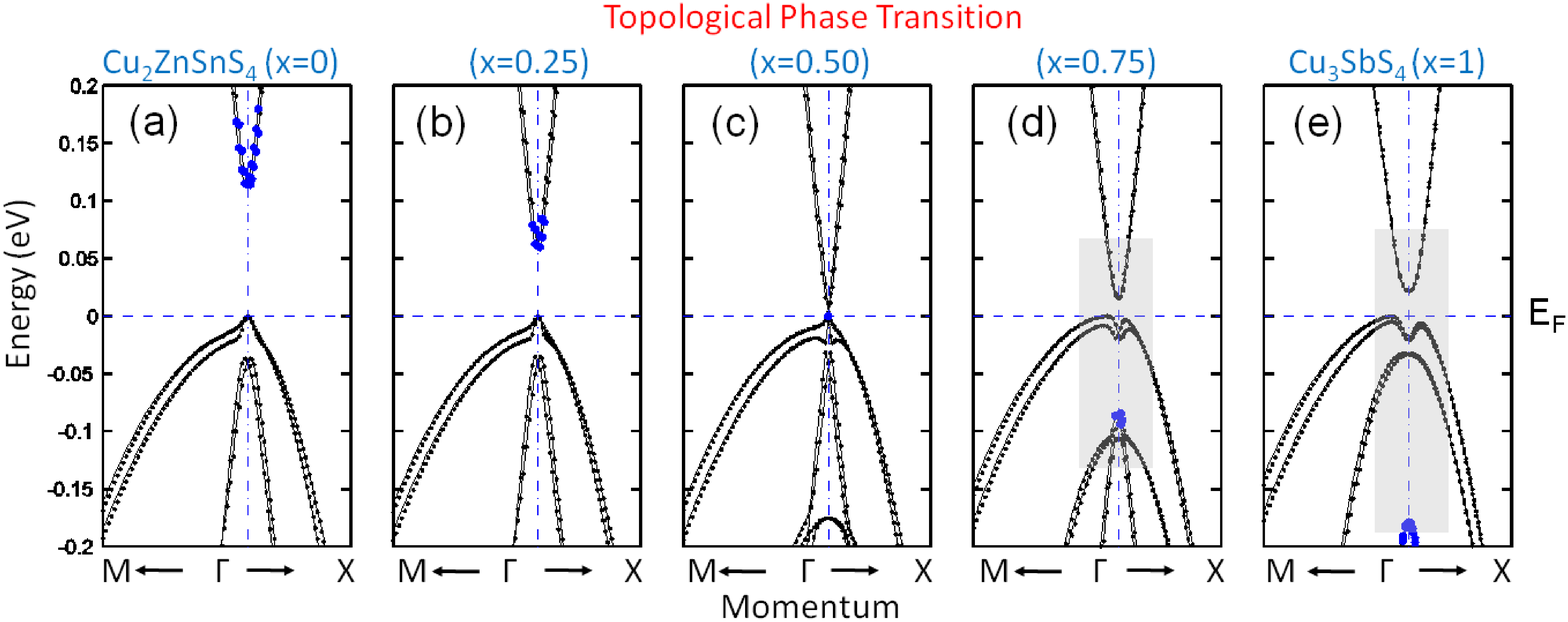}
\caption{\label{fig:bulkbands4} {\bf Topological phase transition between
quaternary chalcogenide and famainite compounds.}
Bulk electronic structure of the compounds at various stages of the
topological phase transition process by tuning the atomic
number $Z$, the lattice constant and the internal displacement of the anion
are shown here. (a) and (b) are the trivial band
insulator, (c) the critical point of the topological phase transition
whereas (d) and (e) are the non-trivial topological insulators.}
\end{figure}
To illustrate this process, we denote the atomic number $Z$ of
various elements
by $Z_{M1}$, $Z_{M2}$, $Z_{M3}$, and $Z_{M4}$.
For our starting element Cu$_{2}$ZnSnS$_{4}$, we have
$Z_{M1}$=29,
$Z_{M2}$=30-$x$,
$Z_{M3}$=50+$x$,
$Z_{M4}$=16,
where $x$ varies between 0 to 1 in the phase transformation process. At each
step of changing the parameter $x$, we also pay attention to the lattice
constant and the internal displacements of the anion by keeping the same
ratio of changes corresponding to the change in $x$ of atomic number Zs from
Cu$_{2}$ZnSnS$_{4}$ to Cu$_3$SbS$_4$. Note that variation of the atomic
number $Z$s can also be considered as a doping effect in the virtual crystal
approximation while the total nuclear charge is neutral during the whole
process.

We monitor the band structure along high symmetry path
$M(\pi,\pi)\rightarrow\Gamma(0,0)\rightarrow X(\pi,0)$ at every step of the
phase transition
process from Figs.~4(a)-(e). While the overall band-structure remains very
much the same throughout the process, the band characters and the band gap
vary dramatically. In Fig.~4(b), on increasing $x$ by $0.25$, the
size of
band gap at $\Gamma$ starts to shrink compared to the parent compound
in Fig.~4(a), although the $s-$character continues to dominate in the
conduction band (blue dots). The critical point of the phase transition is
reached
when the value of $x$ is increased to $0.5$ in Fig.~4(c), at which
point the
bottom of the conduction band and the top of the valence band touch at a
single
momentum point at $\Gamma$ leading to a critical topological point. A full
band inversion is achieved in Fig.~4(d) which recreates an insulating band
gap but topologically non-trivial in nature and the $s-$like orbital
character shifts much below the valence band. Finally, in Fig.~4(e), when
$x$
equals $1$, Cu$_3$SbS$_4$ is a non-trivial topological insulator.

\section{Conclusion}
We have shown via first-principle calculations that the ternary famatinite
and quaternary chalogenide compounds are non-trivial
topological insulators in their pristine phase. The involvement of partially
filled $d-$ and $f-$ electrons in these compounds naturally
opens up the possibility of incorporating
magnetism and superconductivity within the topological order.
The large-scale tunability available in the quaternary compounds adds
versatility in using topolgical insulators in multifunctional
spin-polarized quantum and optical information
procesing applications. Notably, the copper-based quaternary chalcogenides
are known to have non-linear optoelectronics and thermodynamics
applications.\cite{Chenquaternary}.

During the final stages of our work, we became aware of the work of
Su-Huai Wei {\it et al.}\cite{wei2114} who have also proposed that
some of the multinary chalcogenides are potential candidates to be non-trivial
topological insulators.

\textbf{Acknowledgements:}
The work at Northeastern and Princeton is supported by
the Division of Materials Science and Engineering, Basic
Energy Sciences, U.S. Department of Energy Grants No.DE-FG02-07ER46352,
No. DE-FG-02-05ER46200, and No. AC03-76SF00098, and benefited from the
allocation of supercomputer time at NERSC and Northeastern University's
Advanced Scientific Computation Center ASCC.


\section*{References}
\begin{thebibliography}{99}
\bibitem{review}J E Moore
2010 \textit{Nature} \textbf{464} 194-198

\bibitem{reviewMZCL}M Z Hasan and C L Kane 
2010 \textit{Rev. Mod. Phys.} \textbf{82} 3045

\bibitem{Kane1st}Kane C L and Mele E J
2005 \textit{Phys. Rev. Lett.} \textbf{95} 146802

\bibitem{FuKane}Fu L and Kane C L
2007 {\it Phys. Rev. B} \textbf{76} 045302

\bibitem{DavidNat1}Hsieh D, Qian D, Wray L, Xia Y, Hor Y S, Cava R J and
Hasan M Z
2008 {\it Nature} \textbf{452} 970-974

\bibitem{KonigSci}Knig M, Wiedmann S, Brne C, Roth A, Buhmann H, Molenkamp
L W, Qi X.-L and Zhang S.-C
2007 {\it Science} \textbf{318} 766-770

\bibitem{DavidTunable}Hsieh D \textit{et al.}
2009 {\it Nature} \textbf{460} 1101-1105

\bibitem{TIbasic}Kane C L and Mele E J
2007 {\it Phys. Rev. Lett.} \textbf{98} 106803

\bibitem{BernevigSciHgTe}Bernevig B A, Hughes, T L and Zhang S.-C
2006 \textit{Science} \textbf{314} 1757-1761

\bibitem{Roy}Roy, R
2009 {\it Phys. Rev. B} \textbf{79} 195322


\bibitem{MooreandBal}Moore, J E and Balents, L
2007 {\it Phys. Rev. B} \textbf{75} 121306(R)

\bibitem{Majorana}Fu L and  Kane C L
2008 {\it Phys. Rev. Lett.} \textbf{100} 096407

\bibitem{ZhangDyon}Qi X.-L, Hughes T L and Zhang S.-C
2008 {\it Phys. Rev. B} \textbf{78} 195424

\bibitem{KaneSCproximity}Teo J C Y and Kane C L
2010 {\it Phys. Rev. Lett.} \textbf{104} 046401

\bibitem{KaneDevice}Fu L and  Kane C L
2009 {\it Phys. Rev. Lett.} \textbf{102} 216403

\bibitem{cenke}Xu C
2010 {\it Phys. Rev. B} \textbf{81} 054403

\bibitem{palee}Law K T and Ng T K
2009 {\it Phys. Rev. Lett.} \textbf{103} 237001

\bibitem{dhlee}Lee D.-H
2009 {\it Phys. Rev. Lett.} \textbf{103} 196804

\bibitem{DavidScience}Hsieh D \textit{et al.}
2009 {\it Science} \textbf{323} 919-922

\bibitem{MatthewNatPhys}Xia Y \textit{et al.}
2009 {\it Nat. Phys.} \textbf{5} 398-402

\bibitem{Noh}Noh H.-J, Koh H, Oh S.-J, Park J.-H, Kim H.-D, Rameau J D,
Valla T, Kidd T E, Johnson P D, Hu Y and Li Q
2008 {\it Europhys. Lett.} \textbf{81} 57006

\bibitem{ZhangPred}Zhang H \textit{et al.}
Zhang H, Liu C.-X, Qi X.-L, Dai X, Zhong F and Zhang S.-C
2009 {\it Nat. Phys.} \textbf{5} 438-442

\bibitem{ChenBiTe}Chen Y L \textit{et al.}
2009 {\it Science} \textbf{325} 178-181

\bibitem{BiTeSbTe}Hsieh D \textit{et al.}
2009 {\it Phys. Rev. Lett.} \textbf{103} 146401

\bibitem{Roushan}Roushan P, Seo J, Parker C V, Hor Y S, Hsieh D, Qian D,
Richardella A, Hasan M Z, Cava R J and Yazdani A
2009 {\it Nature} \textbf{460} 1106-1109

\bibitem{WrayCuBiSe}Wray L, Xu S, Xiong J, Xia Y, Qian D, Lin H, Bansil A,
Hor Y, Cava R J, Hasan M Z,
2010 {\it Nat. Phys.} \textbf{6} 855–859

\bibitem{heuslerhasan}Lin H, Wray L A, Xia Y, Xu S, Jia S, Cava R J, Bansil
A and Hasan M Z
2010 {\it Nature Mater.} \textbf{9} 546-549

\bibitem{TlBiTe2}Lin H, Markiewicz  R S, Wray  L A, Fu L, Hasan M Z and
Bansil A
2010 {\it Phys. Rev. Lett.} \textbf{105} 036404

\bibitem{Liumagimp}Liu Q, Liu C.-X, Xu C, Qi X.-L, and Zhang S.-C
2009 {\it Phys. Rev. Lett.} {\bf 102} 156603

\bibitem{Hormagimp}Hor, Y. S. {\it et al.}
2010 {\it Phys. Rev. B} {\bf 81} 195203

\bibitem{Yumagimp}Yu R, Zhang W, Zhang H-.J, Zhang S.-C, Dai X and Fang Z
2010 {\it Science} {\bf 329}, 61-64

\bibitem{Akhmerov}Akhmerov A R, Nilsson J and Beenakker C W J
2009 {\it Phys. Rev. Lett.} {\bf 102} 216404

\bibitem{Tanaka}Tanaka Y, Yokoyama T and Nagaosa N
2009 {\it Phys. Rev. Lett.} {\bf 103} 107002

\bibitem{Satotl}Sato T, Segawa K, Guo H, Sugawara K, Souma S, Takahashi T,
and Ando Y
2010 {\it Phys. Rev. Lett.} {\bf 105} 136802

\bibitem{Chentl}Chen Y L {\it et al.}
2010 {\it Phys. Rev. Lett.} {\bf 105} 266401

\bibitem{Chenquaternary}Chen S, Gong X G, Walsh A and Wei S.-H
2009 {\it Phys. Rev. B} {\bf 79} 165211

\bibitem{Liter1}Olekseyuk I D, Piskach L V and Sysa L V
1996 {\it Russ. J. Inorg. Chem.} {\bf 41} 1356-1358

\bibitem{Liter2}Schfer W
1974 {\it Mater. Res. Bull.} {\bf 9} 645-654

\bibitem{Liter3}Olekseyuk I D, Gulay L D, Dudchak I
V, Piskach L V, Parasyuk O V and Marchuk O V
2002 {\it J. Alloys Compd.} {\bf 340} 141-145

\bibitem{Liter4}Parasyuk O V, Parasyuk O V, Gulay L D,
Romanyuk Y E and Olekseyuk I D
2002 {\it J. Alloys Compd.} {\bf 334} 143-146

\bibitem{Liter5}Olekseyuk I D, Marchuk O V, Gulay L D
and Zhbankov O Y
2005 {\it J. Alloys Compd.} {\bf 398} 80-84

\bibitem{Liter6}Kabalov Y M, Evstigneeva T L and
Spiridonov F M
1998 {\it Kristallografiya} {\bf 43} 21-25

\bibitem{Liter7}Olekseyuk I D, Gulay L D, Dudchak I
V, Piskach L V, Parasyuk O V and Marchuk O V
2002 {\it J. Alloys Compd.} {\bf 340} 141-145

\bibitem{Liter8}Parasyuk O V, Gulay L D, Romanyuk Y E
and Piskach L V
2001 {\it J. Alloys Compd.} {\bf 329} 202-207

\bibitem{Liter9}Parasyuk O V, Olekseyuk I D and
Piskach L V
2005 {\it J. Alloys Compd.} {\bf 397} 169-172

\bibitem{Liter10}Guen L and Glaunsinger W S
1980 {\it J. Solid State Chem.} {\bf 35} 10-21

\bibitem{Liter11}Olekseyuk I D, Gulay L D, Dudchak I
V, Piskach L V, Parasyuk O V, Marchuk O V
2002 {\it J. Alloys Compd.} {\bf 340} 141-145

\bibitem{Liter12}Gulay L D, Romanyuk Y E and Parasyuk O V
2002 {\it J. Alloys Compd.} {\bf 347} 193-197

\bibitem{Liter13}Parasyuk O V, Piskach L V, Romanyuk Y
E, Olekseyuk I D, Zaremba V I and Pekhnyo V I
2002 {\it J. Alloys Compd.} {\bf 397} 85-94

\bibitem{Liter14}Pfitzner A
2002 {\it Z. Kristallogr} {\bf 217}, 51-54

\bibitem{Liter15}Pfitzner A
1994 {\it Z. Kristallogr} {\bf 209}, 685

\bibitem{wien2k}Blaha, P.
Schwarz, K., Madsen, G. K. H., Kvasnicka, D. $\&$ Luitz, J.
\textit{WIEN2k, An Augmented Plane Wave Plus Local Orbitals Program for
Calculating Crystal Properties.}
(Karlheinz Schwarz, Techn. University Wien, Austria, 2001).

\bibitem{PBE96} Perdew J P, Burke K and Ernzerhof M
1996 {\it Phys. Rev. Lett.} \textbf{77} 3865-3868

\bibitem{wei2114}Chen S, Gong X G, Duan C.-G, Zhu Z.-Q, Chu J.-H, Walsh A,
Yao Y.-G, Ma J, Wei S.-H
2011 Preprint at $\langle$http://arxiv.org/abs/1104.0353$\rangle$

\end{thebibliography}

\end{document}